\documentclass[aps,prb,amsmath,twocolumn,amssymb,titlepage,]{revtex4-2}

\usepackage{graphicx}
\usepackage{nicefrac}
\usepackage{amsfonts}
\usepackage{xcolor}
\usepackage{amssymb}
\usepackage{amsmath}

\usepackage{subfigure}
\usepackage{multirow} 
\usepackage{tabularx} 
\usepackage{array}

\usepackage{units}

\usepackage{tensor} 
\usepackage{braket}

\usepackage{bm}
\usepackage{hyperref}

\begin{document}
\title{Strain Coupled Domains in BaTiO$_3$(111)-CoFeB Heterostructures}
\author{R. G. Hunt}
\affiliation{School of Physics and Astronomy \\
            University of Leeds, LS2 9JT, United Kingdom}
\author{K. J. A.  Franke}
\affiliation{School of Physics and Astronomy  \\
            University of Leeds, LS2 9JT, United Kingdom}
\author{P. M. Shepley}
\affiliation{School of Physics and Astronomy  \\
            University of Leeds, LS2 9JT, United Kingdom}
      
\author{T. A. Moore}
\affiliation{School of Physics and Astronomy  \\
            University of Leeds, LS2 9JT, United Kingdom}            

\begin{abstract}

Domain pattern transfer from ferroelectric to ferromagnetic materials is a critical step for the electric field control of magnetism and has the potential to provide new schemes for low-power data storage and computing devices.  Here we investigate domain coupling in BaTiO$_3$(111)/CoFeB heterostructures by direct imaging in a wide-field Kerr microscope.  The magnetic easy axis is found to locally change direction as a result of the underlying ferroelectric domains and their polarisation.  By plotting the remanent magnetisation as a function of angle in the plane of the CoFeB layer, we find that the magnetic easy axes in adjacent domains are angled at 60$^\circ$ or 120$^\circ$, corresponding to the angle of rotation of the polarisation from one ferroelectric domain to the next, and that the magnetic domain walls may be charged or uncharged depending on the magnetic field history. Micromagnetic simulations show that the properties of the domain walls vary depending on the magnetoelastic easy axis configuration and the charged or uncharged nature of the wall. The configuration where the easy axis alternates by 60$^\circ$ and a charged wall is initialised exhibits the largest change in domain wall width from 192 nm to 119 nm as a function of in-plane magnetic field.  Domain wall width tuning provides an additional degree of freedom for devices that seek to manipulate magnetic domain walls using strain coupling to ferroelectrics.

\end{abstract}

\maketitle

\section{Introduction}

Multiferroics are an attractive set of materials for future device engineering owing to the ability to control magnetic properties purely through the application of voltages, offering a low power option to integrate into existing magnet-based devices\cite{Guo2021}. Ideal single-phase multiferroics developed up to now, in which the material has direct coupling between ferroelectric and ferromagnetic order, exist only at cryogenic temperatures, limiting the applicability of these materials in devices\cite{Lu2019}. For room temperature purposes, the most attractive options currently available include BiFeO$_3$\cite{Sando2014}, a single phase ferroelectric-antiferromagnet system, and multiferroic heterostructures that couple two materials with different ferroic ordering at the interface through mechanisms such as strain transfer\cite{Spaldin2019} or interfacial magnetoelectric coupling in epitaxially grown magnetic films\cite{Li2022,Radaelli2014}. Significant work has already been carried out on the strain coupling of ferroelectric substrates and ferromagnetic thin films with a great deal of success showing control of magnetic domain wall position and width, and magnetic anisotropy via application of magnetic fields\cite{LopezGonzalez2017}, electric fields\cite{Franke2015,Lahtinen2012}, and optical pulses\cite{Shelukhin2020}.

Strain coupled heterostructures present an interesting system in which the magnetization is strongly coupled to the ferroelectric domains and there is a local level of control over the magnetization. In a magnetostrictive material a strain-dependent anisotropy is introduced as $K_{me} \propto \lambda \sigma $ where $K_{me}$ is the magnetostrictive anisotropy induced by inverse magnetostriction, $\lambda$ is the magnetostriction and $\sigma$ is the applied stress. In these heterostructures $\sigma$ is the result of interfacial strain from the substrate to the film and will change orientation dependent upon the polarization direction of the local ferroelectric domains. The rotation of these imprinted magnetoelastic anisotropies leads to reduction in the angle by which the magnetization rotates from one domain to the other, deviating from the 180$^{\circ}$ rotation of a typical Bloch domain wall leading to either head-to-tail domain walls that are similar in nature to Bloch walls or head-to-head charged magnetic domain walls\cite{Franke2012} in which the magnetization points into the domain wall from both sides. Devices depending on the ability to freely switch between these domain wall states have already been proposed\cite{Hamalainen2018, LopezGonzalez2016} taking advantage of the reliability at which domain walls form at ferroelectric domain boundaries and the ability to initialize a charged or uncharged state.

\begin{figure}
\includegraphics[width=\columnwidth]{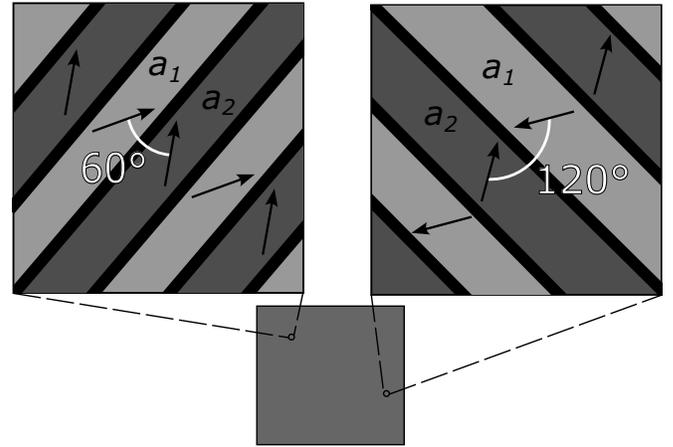}
\caption{\label{fig:schematic} Schematic diagram of ferroelectric domains within a BTO(111) substrate. Both 60$^{\circ}$ and 120$^{\circ}$ domain walls can exist as shown in the two blown-up regions of the substrate. Arrows indicate the direction of ferroelectric polarization.}
\end{figure}

In this work we investigate the domain pattern transfer of a BaTiO$_3$ (BTO) (111)-oriented substrate onto a CoFeB thin film. We present our results in the room-temperature tetragonal phase of the BTO. In the (111) plane the projection of the electric polarization has equivalent directions angled at 60$^\circ$ or 120$^\circ$ to each other. To minimize the static dipolar energy the ferroelectric substrate forms domains in which the polarization alternates between equivalent crystal directions. At domain wall boundaries the polarization then rotates through 60$^{\circ}$ either through the domain wall boundary resulting in a 60$^{\circ}$ domain configuration or normal to the domain wall boundary resulting in a 120$^{\circ}$ domain configuration. This is sketched in Fig. \ref{fig:schematic} where we highlight the existence of both of these types of domains at different locations in a sample. The polarization in BTO is the result of lattice elongations andfor a sufficiently soft, magnetostrictive material, such as a sputtered CoFeB thin film, in which the strain is efficiently transferred, an in-plane easy axis will be imprinted into the ferromagnet with a local dependence upon the domain structure of the substrate.

Compared with previous work focusing on BTO(100) heterostructures\cite{Casiraghi2015, Franke2012, Lahtinen2012}, the (111) orientation introduces an additional in-plane domain configuration through which we can potentially control the magnetic properties of a ferromagnetic film, in particular the properties at the domain wall which could be useful for domain wall based applications.

\begin{figure*}
\includegraphics[width=1.7\columnwidth]{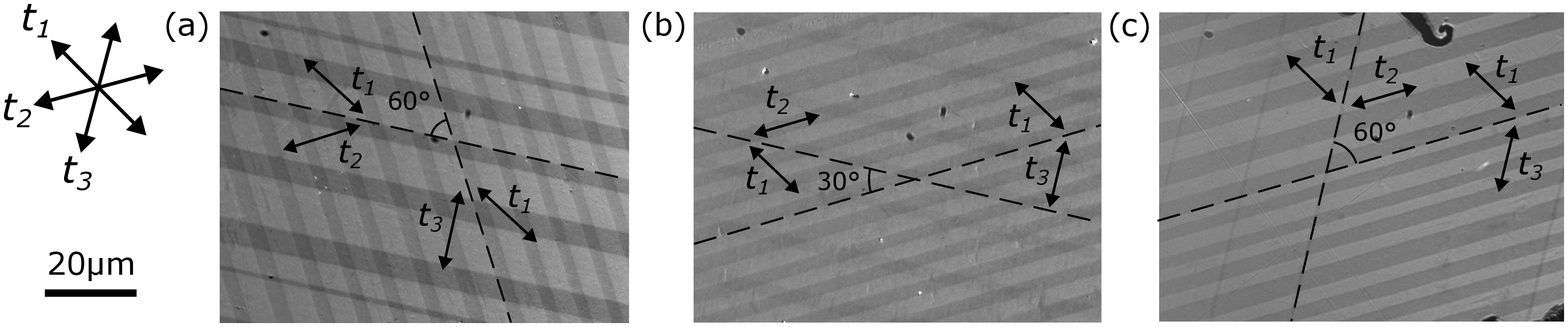}
\caption{\label{fig:fe_domains} Polarised light microscopy of ferroelectric domains in BTO(111) in three distinct regions, A, B and C, illustrating different orientations of ferroelectric domain walls. Dashed lines indicate the ferroelectric domain walls, and double headed arrows represent possible directions of lattice elongation.}
\end{figure*}

\section{Experimental Results}

Samples were fabricated in the Royce deposition system\cite{Royce} via DC magnetron sputtering in an Ar pressure of 4.3$\times10^{-3}$ mbar . A thin film of Co$_{40}$Fe$_{40}$B$_{20}$ with a thickness of 20 nm is deposited at 300$^{\circ}$C onto a BTO(111) 5$\times$5$\times$0.5mm substrate, above the ferroelectric Curie temperature of the BTO, and is subsequently cooled through this transition to establish strain transfer from the ferroelectric domains to the ferromagnetic film. A capping layer of 5 nm Pt is deposited at room temperature.

Magnetic measurements are made using a wide field Kerr microscope with spatial resolution 0.2 $\mu$m in which it is possible to select a single region of a sample for surface-sensitive magnetic measurements and to then obtain hysteresis curves from individual domains corresponding to different regions of ferroelectric coupling.

\begin{figure}
\includegraphics[width=\columnwidth]{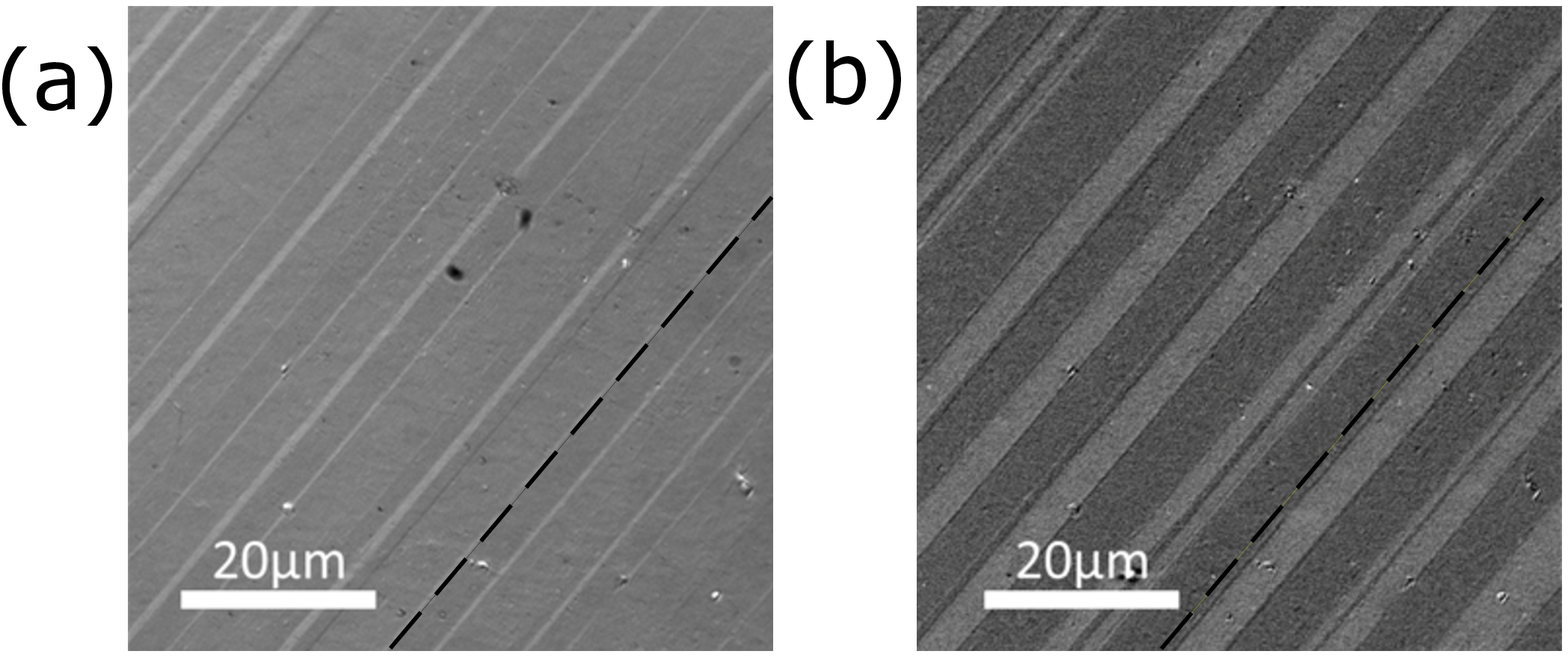}
\caption{\label{fig:coupling} Coupling between a) ferroelectric and b) ferromagnetic domains in BTO(111)/CoFeB. Changes in magnetic contrast line up exactly with ferroelectric domain walls. The dashed line indicates the position of the same domain wall in each image.}
\end{figure}

Initial imaging confirms the presence of ferroelectric domains at the surface of the substrate visible through the magnetic film and capping layer. These domains are observed using a polarized light source and contrast is obtained as a result of the difference in refractive indices of the different ferroelectric domains. This is an intrinsic property of the crystal structure of the BaTiO$_3$ in the tetragonal phase and has been well-documented in the literature\cite{Wada1999, Forsbergh1949}. Magnetic contributions to the contrast are removed by using pairs of LEDs on opposing sides to result in no net contrast from the in-plane magnetic domains.

From our ferroelectric imaging we find that both configurations shown in Fig. \ref{fig:schematic} exist in our substrates. Fig. \ref{fig:fe_domains} shows images from three regions (a), (b) and (c) taken in the same state. Each image shows regions in which it is possible to view two ferroelectric domains that overlap through the depth of the substrate. Represented alongside this are the three directions of lattice elongation, $t_1$, $t_2$, and $t_3$ within the same frame of reference that are offset from each other by 60$^\circ$. The orientation of domain walls is highlighted by dashed lines. 

Given the domain structures that are possible the orientation of the ferroelectric domain walls is constrained such that domain configurations of the same type (60$^\circ$ or 120$^\circ$) will have domain walls oriented at 60$^\circ$ to each other, whereas configurations of opposite type will be have domain walls oriented at 30$^\circ$ or 90$^\circ$ from one to the other. If we make the assumption that in Fig. \ref{fig:fe_domains} (a) that both domains are 60$^\circ$ domains then we know from the angle between domain walls in (b) that (c) must correspond to a 120$^\circ$ configuration, and the opposite would be true if (a) was instead a 120$^\circ$ configuration and we thus show that all ferroelectric configurations and polarizations exist within the substrate.

\begin{figure}
\includegraphics[width=0.9\columnwidth]{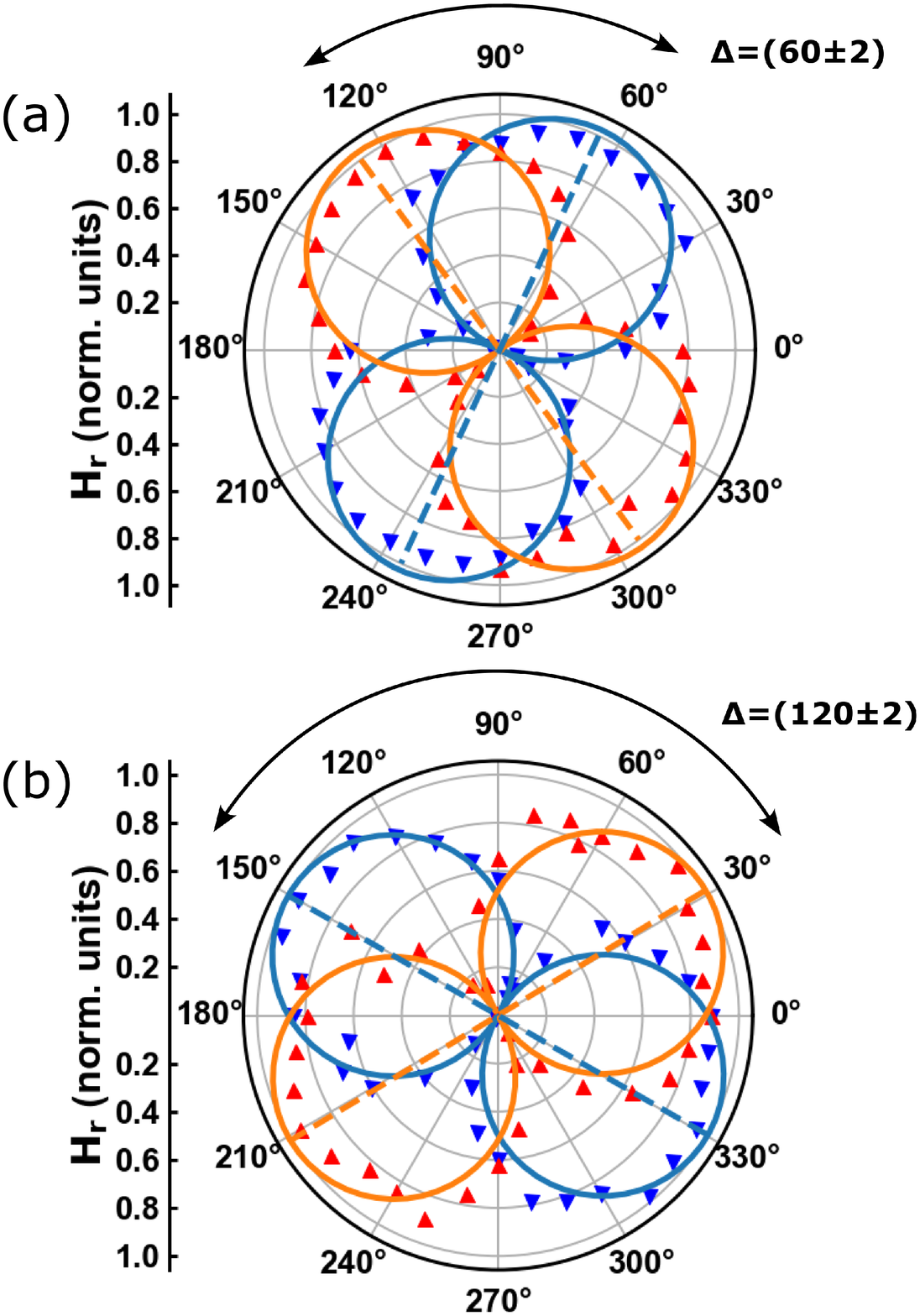}
\caption{\label{fig:anisotropy_plot}Remanent magnetization taken from hysteresis loops in individual magnetic domains for a) 60$^{\circ}$ and b) 120$^{\circ}$ configurations, with the orientation of the easy axes marked by dashed lines. Red and blue triangles represent data adjacent magnetic stripes from which data is obtained. Red and blue triangles represent adjacent stripes from which data is obtained. $\Delta$ is the change in angle through the domain wall boundary, here oriented at 90$^\circ$.}
\end{figure}

\begin{figure}
\includegraphics[width=0.9\columnwidth]{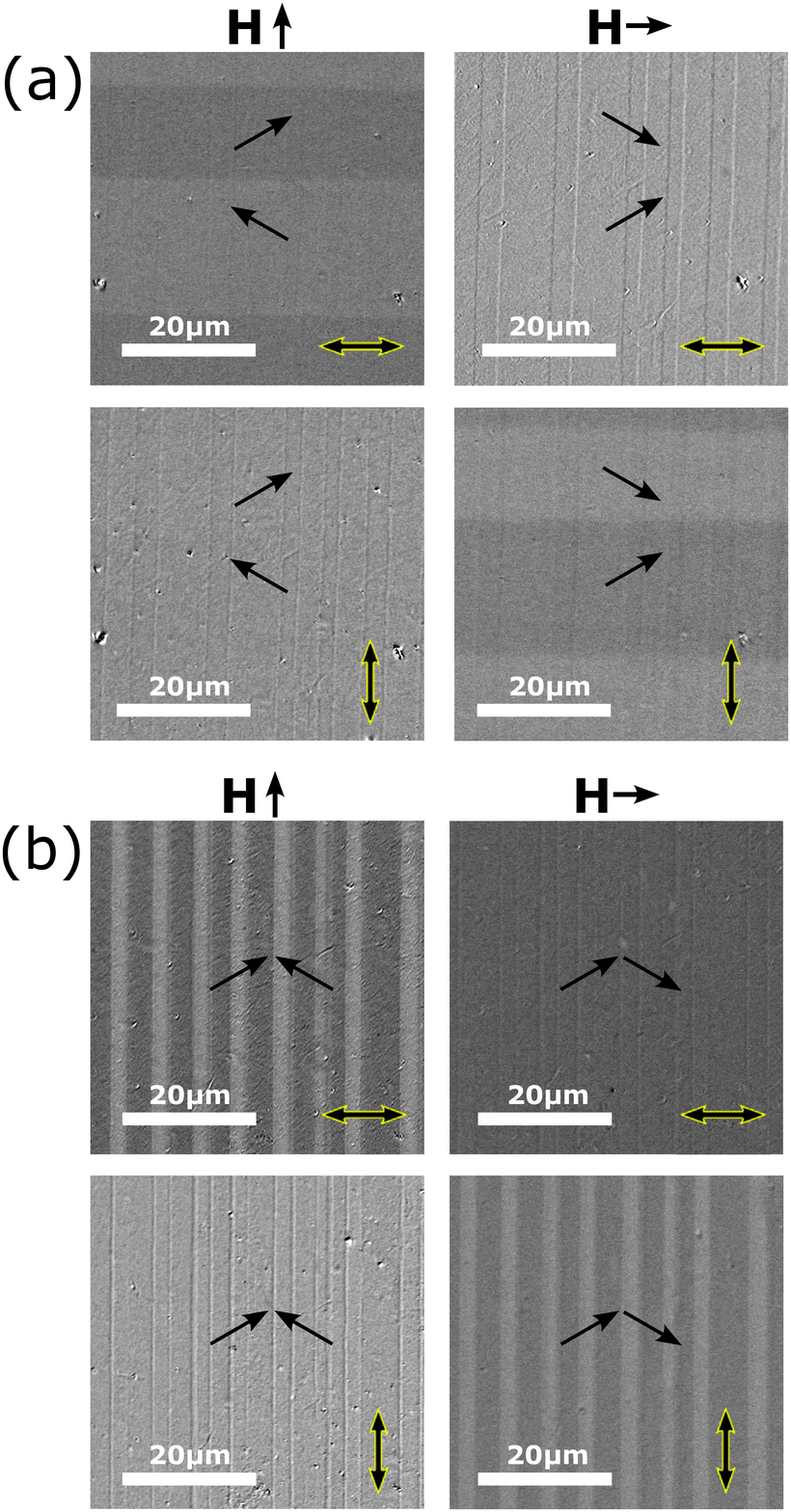}
\caption{\label{fig:charged_uncharged} All four field-contrast combinations for a) 60$^\circ$ and b) 120$^\circ$ domain patterns. Black arrows indicate the local direction of magnetization, black and yellow double headed arrows show the direction of magnetic contrast and the direction of applied field is shown at the top of the column.}
\end{figure}

Magnetic contrast is obtained using a subtraction technique to amplify the changes in light due to Kerr rotation and we confirm the strain coupling between ferroelectric and ferromagnetic domains in Fig. \ref{fig:coupling}. By comparing the ferroelectric and ferromagnetic images we observe that the positions of the domain walls match exactly, one of which is highlighted by the dashed line in both images representing the same domain wall.

\begin{figure*}
\includegraphics[width=2\columnwidth]{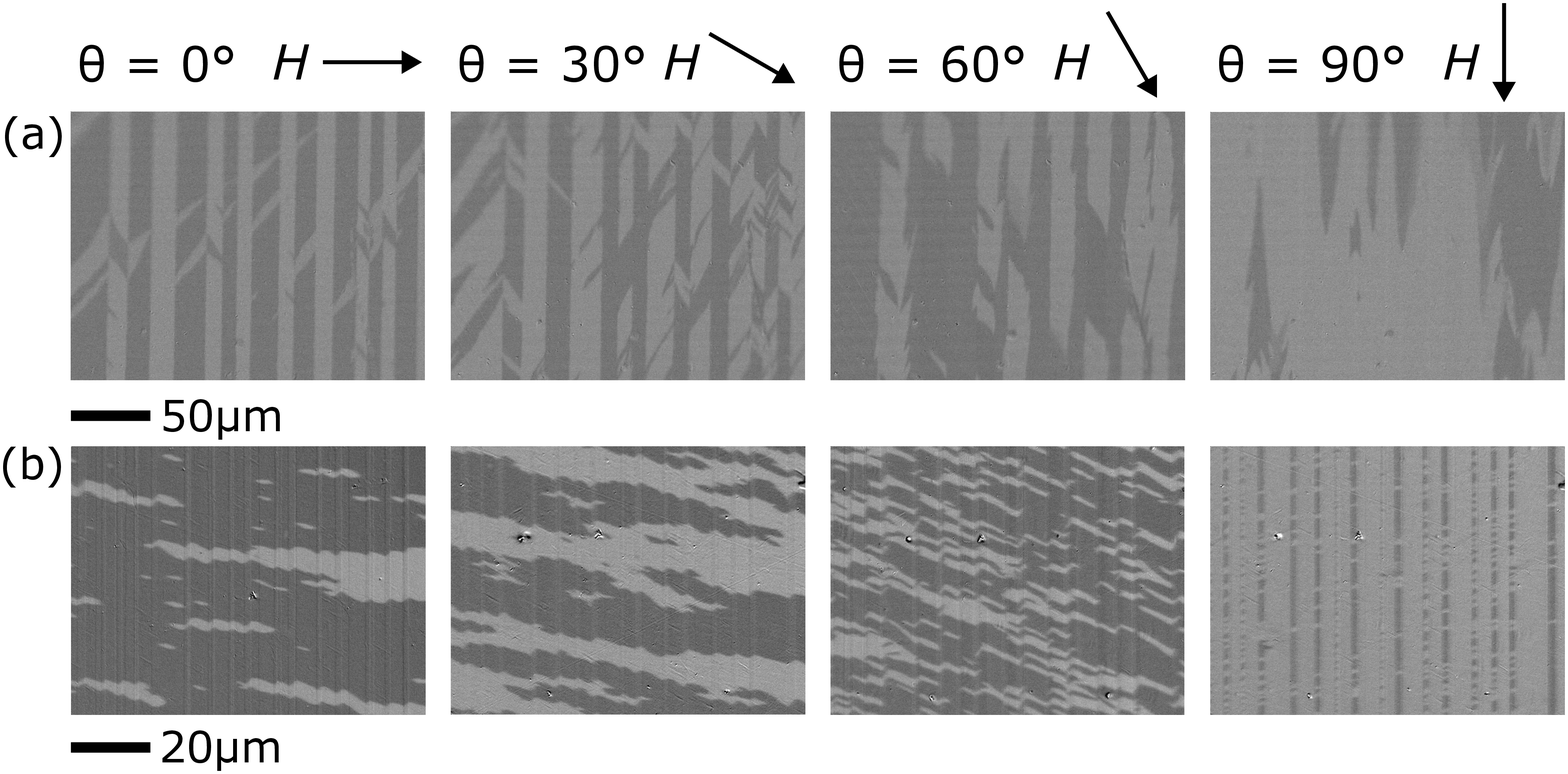}
\caption{\label{fig:domain_rotation} Magnetization reversal for a) 60$^\circ$ and b) 120$^\circ$ domain configurations as a function of field angle. Direction of the magnetic field is indicated above the images and scale bars are indicated below. Each of the eight reversal processes defined by domain wall configuration and field angle is represented by a single image.}
\end{figure*}

The domains are further investigated by taking polar plots of the remanent magnetization which acts as a representation of how close to the easy axis each direction is, with a remanence close to 1 (in normalized units) representing a purely easy-axis aligned hysteresis loop and a remanence close to 0 being the hard axis. Angles are measured relative to the ferroelectric domain wall so that comparisons can be more easily drawn between different configurations. In Fig.\ref{fig:anisotropy_plot} the results are shown for two different regions of the film corresponding to the 60$^{\circ}$ and 120$^{\circ}$ configurations. The data is fitted to a $R\cos^{2}(\theta-\theta_0)$ function where $R$ is an arbitrary scaling parameter and $\theta_0$ is the offset from zero that represents the orientation of the in-plane easy axis from the inverse magnetostriction. For both configurations the angle between the easy axes in adjacent domains, $\Delta$, agrees well with a rotation of 60$^{\circ}$ and 120$^{\circ}$ for the respective configurations.  From our hard axis loops we find the in-plane anisotropy field and calculate the corresponding anisotropy constant from $H_{k} = \frac{2K_{eff}}{\mu_{0}M_{s}}$ to obtain a value of $K_{me} = 3.0\times10^4$ J/m$^3$ for the 60$^{\circ}$ domain configuration and $K_{me} = 2.5\times10^4$ J/m$^3$ for the 120$^{\circ}$. In our samples we find that the 120$^{\circ}$ domain regions are generally much smaller by up to an order of magnitude, typically in the region of microns, while the 60$^{\circ}$ regions can be 10s of microns in scale. This could explain the difference in the magnitudes of $K_{me}$ as it has been shown previously\cite{Casiraghi2015} to decrease with reduced ferroelectric domain width, with a more pronounced effect in thinner films.

Using the previously identified regions we apply a 40mT magnetic field parallel and perpendicular to the ferroelectric stripe direction and image the sample with longitudinal and transverse contrast\cite{Soldatov2017} to obtain images corresponding to the charged and uncharged domain wall states. Fig. \ref{fig:charged_uncharged} shows the charged and uncharged 60$^{\circ}$ (Fig. \ref{fig:charged_uncharged}a) and 120$^{\circ}$ (Fig. \ref{fig:charged_uncharged}b) configurations. For both configurations the application of a magnetic field along the length of the stripe produces a magnetically uncharged domain wall type in which the magnetization smoothly rotates from one domain to the other, and application perpendicular to the stripe produces a magnetically charged domain wall. The presence of both is in agreement with similar experiments in BTO(100) substrates where the charged and uncharged domains form in the same alignment as in these experiments\cite{Franke2014}. The charged and uncharged domain walls in the 60$^{\circ}$ and 120$^{\circ}$ configurations are distinct and distinguished by the change in magnetization angle from one domain to the other. Based on the directions of the easy axes in adjacent stripes and the contrast shown in Fig. \ref{fig:charged_uncharged} we can extrapolate the spin rotation through the magnetic domain wall for each domain wall type. In the 60$^{\circ}$ configuration the charged domain wall has a spin rotation of 60$^{\circ}$ and the uncharged has a spin rotation of 120$^{\circ}$, and the opposite is true for the 120$^{\circ}$ configuration. This results in a total of four unique domain wall structures in these BTO(111)-based heterostructures as shown in Fig. \ref{fig:charged_uncharged}.

\begin{figure}
\includegraphics[width=\columnwidth]{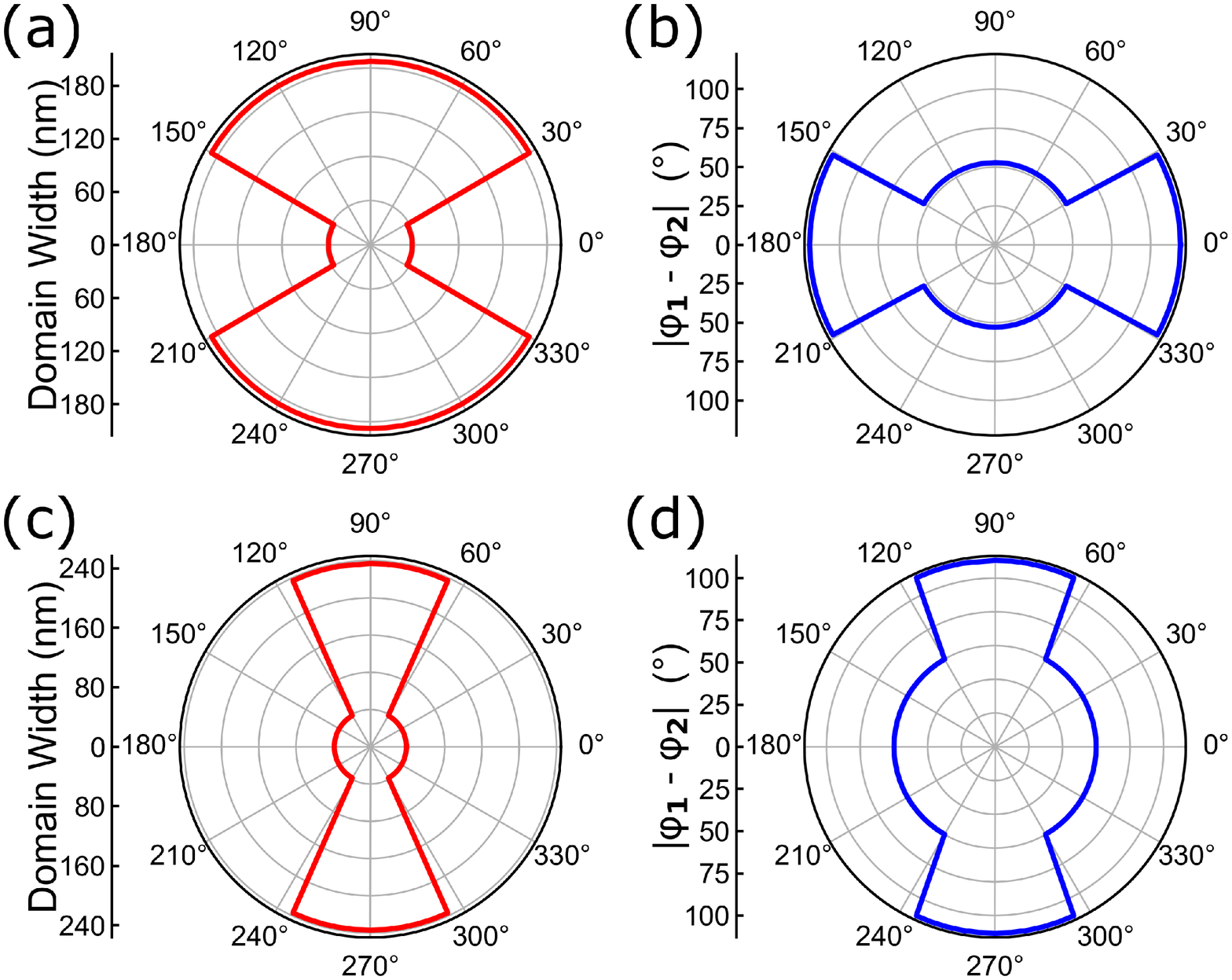}
\caption{\label{fig:polar_simulation} Micromagnetic simulations of the domain width (a), (c) and spin rotation (b), (d) for 60$^{\circ}$ and 120$^{\circ}$ configurations respectively as a magnetic field is rotated with respect to the ferroelectric domain wall.}
\end{figure}

The type of domain wall was found to have an impact on the propagation of magnetic domains in both of these regions. The domain propagation follows the elongation axis set by the ferroelectric domains which makes for visually distinct domain patterns dependent upon the ferroelectric domain configuration. For a 60$^{\circ}$ domain wall the acute angle results in arrow-head domains and for 120$^{\circ}$ domain walls the magnetic domains propagate over long distances in a staircase pattern, both of which can be seen in Fig. \ref{fig:domain_rotation} a) and b) respectively for the $\theta=0^{\circ}$ case. The angle of the magnetic domain wall to the ferroelectric domain wall agrees well with the angles between the easy axes found in Fig. \ref{fig:anisotropy_plot}, although there is some deviation as a result of defects in the magnetic film, magnetic domains joining, and various other effects that cause the magnetic domains to bend and change as they propagate particularly over long distances in the very wide ferroelectric domains where this deviation is more pronounced. 

The reversal domain patterns have a strong dependence on the angle of applied magnetic field with respect to the domain wall. To show this we saturate the film and nucleate domains with a field of 17mT, close to the coercivity, for different field directions. The change in domain formation pattern is summarized in Fig. \ref{fig:domain_rotation} with the field angle shown for each domain image produced. If the field is applied perpendicular to the ferroelectric domains, the domains form in a co-operative manner in which domains can easily propagate from one stripe to the next. This gives rise to the arrow-head and staircase domains previously mentioned. As the angle increases and approaches the direction of lattice elongation the density of domains nucleated increases and eventually when the angle exceeds the magnetoelastic easy axis of one of the ferroelectric domains then domains are nucleated preferably in one set of ferroelectric stripes. In this state the domains nucleate along the length of one set of stripes first and then the other set of stripes switch. Qualitatively, this agrees with results obtained previously for BTO(100)/CoFeB heterostructures\cite{Casiraghi2015} where it was found that applying a magnetic field parallel to the domain wall results in one set of stripes switching preferably before the other set, and applying the field perpendicular showed no such behaviour. We can see that the transition agrees well with the results shown in Fig. \ref{fig:anisotropy_plot} with the transition for the 60$^{\circ}$ state in Fig. \ref{fig:domain_rotation} a) occurring at 30$^{\circ}$ and the transition for the 120$^{\circ}$ state in Fig. \ref{fig:domain_rotation} b) occurring at 60$^{\circ}$.

\section{Micromagnetic Simulations}

We now demonstrate several further properties of these films using the results obtained from our experiments to simulate the system in a micromagnetic framework for which we choose to use MuMax3\cite{Vansteenkiste2014}. This has previously been used to simulate very similar heterostructures and has been shown to be in good agreement with experimental results. For our simulations we use micromagnetic parameters of saturation magnetization $M_{sat}$ = $854\times10^3$ A/m, exchange stiffness $A_{ex}$ = 2.1$\times$10$^{-11}$ J/m, magnetoelastic anisotropy $K_{u1}$ = 3$\times$10$^4$ J/m$^{3}$ for the 60$^{\circ}$ configuration and $K_{u1}$=2.5$\times$10$^4$ J/m$^{3}$ for the 120$^{\circ}$ configuration. The values of saturation magnetization and anisotropy strength have been obtained from our experiments while the exchange stiffness is taken from previous work\cite{Franke2014}. The system is divided into two regions with uniaxial anisotropy vectors corresponding to the direction of in-plane lattice elongation in adjacent ferroelectric stripes dependent upon which of the two domain types is being simulated. Within this setup we investigate the change in the spin rotation, with angles taken relative to the orientation of the magnetoelastic easy axis, and domain wall width defined as:

\begin{equation}
    \delta = \int_{-\infty}^{\infty} \cos^{2}(\phi') \,dx \,
\end{equation}

where $\phi'$ is the reduced magnetization angle defined as,

\begin{equation}
    \phi' = \bigg(\phi-\frac{\mid\phi_{\frac{\Delta}{2}} - \phi_{\frac{-\Delta}{2}}\mid}
                {2}\bigg) 
            \frac{180}
                {\mid{\phi_{\frac{\Delta}{2}} - \phi_{-\frac{\Delta}{2}}\mid}}
\end{equation}

and the angles $\phi_{\frac{\Delta}{2}}$ and $\phi_{-\frac{\Delta}{2}}$ are the angles of the magnetization far from the domain wall on either side. 

This method of calculating the domain width has been used previously\cite{Hubert1998,Franke2014} and encompasses the entirety of the spin rotation in a reduced domain wall angle system much better than fitting to a $tanh$ function as would normally be performed for a Bloch type domain wall.

In our simulations we apply a saturating field at a variety of angles and then allow the system to relax in zero field. The domain wall reliably forms at the boundary between adjacent easy axes and it is around this point that we calculate the spin rotation and domain wall profiles. The angle of the applied field is rotated from $\theta=0^{\circ}$ along the normal of the domain wall and the polar plots in Fig. \ref{fig:polar_simulation} are produced. Domains produced in this way reliably relax into only the charged or uncharged state, with the domain width for the charged state being an order of magnitude greater. Surprisingly, the domain width has no angular dependence close to the transition angle which is different for each configuration, but instead sharply transitions from an uncharged domain wall to a charged domain wall when the field aligns with one of the magnetoelastic anisotropy axes. This agrees well with the results produced earlier in Fig. \ref{fig:domain_rotation} in which crossing a magnetoelastic easy axis changes the way in which domains propagate through the film - this is the result of the applied field changing from producing an uncharged or charged domain wall state. 

The spin rotation in both of the charged domain wall states is reduced, from 60$^{\circ}$ or 120$^{\circ}$ to 53$^{\circ}$ and 111$^{\circ}$ for the 60C and 120C states respectively. This is a result of the value of saturation magnetization used in these simulations. For charged domain walls the saturation magnetization has a large impact on the domain width\cite{Franke2021}, and this leads to a suppression of the spin rotation. The spin rotation of charged states is also much more heavily dependent on the thickness of the film than the uncharged state and this will also contribute to the suppression of the rotation\cite{Franke2021a}.

\begin{figure}
\includegraphics[width=\columnwidth]{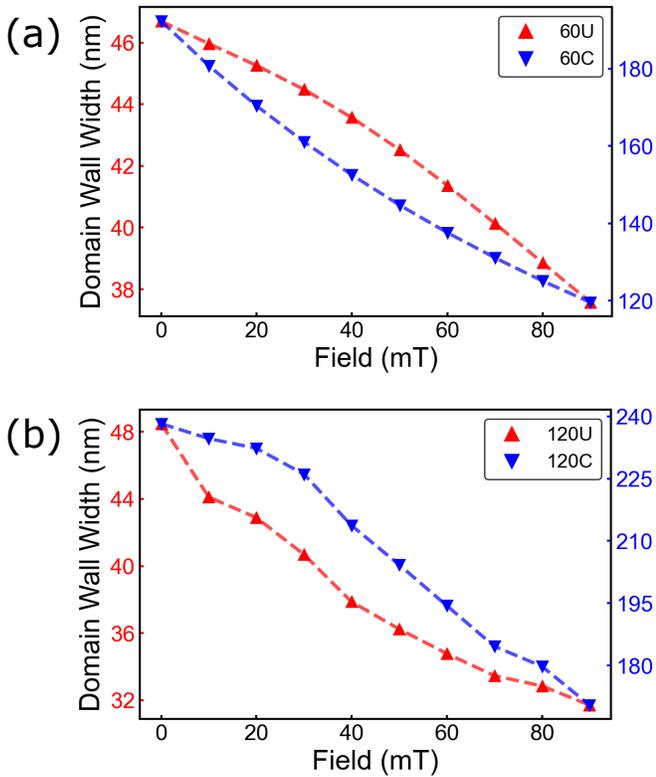}
\caption{\label{fig:field_scaling} Magnetic field dependence of domain width (a) the 60$^{\circ}$ configuration and (b) the 120$^{\circ}$. Left and right axes correspond to the uncharged and charged domain wall widths respectively.}
\end{figure}

We investigate how variable these two parameters are under an applied magnetic field in all four domain wall configurations. For ease of reference we use the convention of the literature in which configurations are defined as the angle between magnetization in adjacent domains and the uncharged/charged nature of the domain wall, 60U, 60C, 120U and 120C. In all instances the spin rotation behaves the same under an applied magnetic field, with an asymptotic drop off in amplitude that corresponds to the gradual realignment of spins from the easy axes to the field direction. The domain width however shows two characteristic behaviours. In Fig. \ref{fig:field_scaling}a the 60C configuration follows the realignment of spins while the 60U domain wall is initially more resistive and mirrors the realignment of spins. In Fig. \ref{fig:field_scaling}b the behaviour is inverted with the 120C wall showing a similar trend to the 60U domain wall, and the 120U and 60C also being similar in character. From this we can understand the behaviour of the domain wall depends on the rotation of magnetization between adjacent domains moreso than the charged or uncharged nature.

The four different configurations have different responses that have important implications for any applications that could make use of domain wall width. The percentage change of domain width as compared to the initial state is greatest in the 60C (38\%) and 120U (35\%) where the spin rotation drops off from 60$^\circ$ and follows the realignment of spins to the field. Conversely in the large angle states the relationship is a mirror of the spin rotation and is much more resistive to the effects of an applied field demonstrating a lower reduction in the domain width for both the 120C (28\%) and 60U (20\%) states. 

\section{Conclusion}

In conclusion we have demonstrated the pattern transfer from ferroelectric stripes in BTO(111) substrates to a CoFeB film and shown that there are four unique magnetic domain wall configurations that can be initialized in these heterostructures corresponding to rotations in the easy axes of either 60$^{\circ}$ or 120$^{\circ}$ and their charged or uncharged variations. These regions behave differently with entirely different domain patterns and switching mechanisms controlled purely by the angle between adjacent magnetoelastic anisotropy axes in adjacent domains.

Using micromagnetic simulations we are able to confirm our observations by calculating the angular dependence of the domain wall width showing that the transition from uncharged to charged wall happens abruptly as the field direction crosses the orientation of the lattice elongations in-plane. We predict that the response of the domain widths for each of these states is unique, with the charged 60$^{\circ}$ state presenting the best case for domain wall width tunability both in terms of absolute change and in terms of the response type with this case exhibiting the most linear change. This presents an additional option for control of devices in which a magnetic domain wall could be the core component.

\section{Acknowledgements}

We acknowledge support from the Henry Royce Institute and funding from the Engineering and Physical Sciences Research Council (EPSRC) Grant No. EP/M000923/1. This project received funding in part from the European Union’s Horizon 2020 research and innovation programme under the Marie Sklodowska-Curie grant agreement No 750147. R. G. H. acknowledges the support of an EPSRC DTA studentship. 

\bibliography{BTO111-CoFeB.bib}

\end{document}